\documentclass[aps,preprint,showpacs,superscriptaddress,groupedaddress]{revtex4}  % for double-spaced preprint
\usepackage{graphicx}  % needed for figures
\usepackage{dcolumn}   % needed for some tables
\usepackage{bm}        % for math
\usepackage{amssymb}   % for math
\newcommand{\bra}[1]{\big\langle \, #1\,\big\vert}
\newcommand{\ket}[1]{\big\vert\, #1\,\big\rangle}
\newcommand{\bracket}[2]{\big\langle \, #1 \big\vert \, #2\,\big\rangle}

\begin{document}

% The following information is for internal review, please remove them for submission
%\widetext
%\leftline{Version 1.2 as of \today}
%\leftline{Primary authors: N. Nikitin, V. Sotnikov, K. Toms}
%\leftline{Submitted to PRA}

\title{A new class of time-dependent Bell inequalities in Wigner form}

\author{N. Nikitin}
\author{V. Sotnikov}
\affiliation{Lomonosov Moscow State University Skobeltsyn Institute of
  Nuclear Physics, Russia}
\author{K. Toms}
\affiliation{Department of Physics and Astronomy, University of New
  Mexico, USA}
\date{\today}

\begin{abstract}
We derive a new class of time-dependent Bell inequalities in Wigner
form under the assumption of locality in the framework of Kolmogorov's
probability theory. We consider violation of the obtained
inequalities for three cases: spin correlations in an external
magnetic field; oscillations of neutral pseudoscalar mesons; 
and decays of a pseudoscalar into a fermion-antifermion pair.
\end{abstract}

 %http://www.aip.org/pacs
\maketitle

%\section{\label{sec:level1}First-level heading}
% sections are not used for PRL papers

\section{Introduction}

One of the fundamental questions that appear in both non-relativistic quantum
mechanics (NRQM) and quantum field theory (QFT) is how the physical properties
of micro-objects are related to the measurement procedure performed
with a macroscopic device. According to the Copenhagen interpretation of
quantum mechanics, one can refer to properties of
micro-objects only if the macroscopic measurement procedure for them is
defined. The maximum amount of information related to a micro-object's
properties is then defined by the number of its ``characteristics''
which can be measured by a given macro-device. In quantum theory these
characteristics are represented by sets of commuting Hermitan
operators.

However it is possible to introduce non-commuting operators
corresponding to physical observables for a given micro-system,
e.g. the operator of spin $s=1/2$ projection onto some non-parallel
directions. In NRQM and QFT this operator is usually defined as
$\vec s = \vec O/2$ \footnote{In NRQM $\vec O = \vec
  \sigma$. In QFT the $s=1/2$ operator may be defined in many
  ways. For studies of Bell inequalities it is suitable to use the
  following definition: $\displaystyle \vec O\, =\,
  -\,\gamma^5\,\vec\gamma\, +\,\gamma^5\,\frac{\vec
    p}{\varepsilon_p}\, +\, \frac{\vec p\,\gamma_5\,
    (\vec\gamma,\,\vec p\, )}{\varepsilon_p\, (\varepsilon_p + m)}$,
  where $\gamma^5\, =\, i \gamma^0 \gamma^1 \gamma^2 \gamma^3$,
  $\varepsilon_p$ = energy, $\vec p$ = momentum, and $m$ = mass of the
  fermion.}.  Cartesian projections of this operator satisfy the
following commutation relations:
$$
[O^i, O^j] = 2i\epsilon^{ijk} O^k, \mathrm{~where~} \epsilon^{123}=+1.
$$
The above operators do not have a common system of eigenvectors, hence
it is impossible to measure simultaneously any two spin projections
onto non-parallel directions. Bohr's principle of complementarity is a
philosophical reflection of this fact, while Heisenberg's uncertainty
principle is a mathematical one. 

Is it possible for characteristics of a macro-system, which are
described by non-commuting operators, to be simultaneously the
elements of physical reality, even in the absence of a macro-device
which is able to measure all of them simultaneously? This
question was first introduced in \cite{Einstein:1935rr}, and N.~Bohr
addressed this in \cite{Bohr:1935af}, presenting another philosophic
view on the problem of measurement in the NRQM. Fifty years ago
J.~Bell suggested an experimental method
\cite{Bell:1964kc,Bell:1964fg,nature510}, which has been further
developed by J.~Clauser, M.~Horne, A~Shimony and R.~Holt
\cite{Clauser:1969ny}, and has been widely discussed since then
\cite{epr-books,Reid:2009zz,Rosset:2014tsa}.

One of the ways to express mathematically the impossibility of
simultaneous measurement of a set of observables with a single
macro-device is to suggest that the joint probability of existence of
the given set is non-negative. This assumption has been used by
J.~Bell in \cite{Bell:1964kc}, although not as an explicit statement,
as was done later by E.~Wigner \cite{wigner}. The inequalities obtained
are now known as Bell inequalities in Wigner form or just
as Wigner inequalities. Calculation or measurement of the probability
in quantum mechanics is a well defined procedure, making the Bell
inequalities in Wigner form more ``natural'' in some sense than the
classic Bell inequalities, which are formulated in terms of
correlators of operators corresponding to two observables. Derivation
of classic Bell inequalities using the non-negative joint probabilities
has been done in \cite{muynck1986}. 

An attempt at a relativistic generalization of the Bell inequalities in
Wigner form may be found in \cite{Nikitin:2009sr}, where various
relativistic corrections were considered for decay of a pseudoscalar
particle into a fermion-antifermion pair. Corrections due to
non-parallel momenta of fermions (e.g. due to soft photon radiation)
were taken into account, as well as corrections due to finite
distance to the spin analyzers. It was shown that these effects have
almost negligible influence upon the classic Wigner inequalities. 
The time dependence, however, was not taken into account in
\cite{Nikitin:2009sr}.

There are well known time-dependent Leggett-Garg inequalities
\cite{Leggett:1985zz}, based on the idea of macroscopic realism,
i.e. that every physical observable has a defined value at every
moment of time, and that the measurement performed does not affect the
subsequent dynamics of the observable. Leggett-Garg inequalities use
the correlation between the values of the same observable at distinct
moments of time. For example, it is possible to examine the correlations
between spin projections for the spin precession in an external magnetic
field \cite{Kofler-2007}. Experimental study of violation of
Leggett-Garg inequalities is closely related to the idea of ``weak
measurement'' \cite{Aharonov:1988xu,korotkov-2006};
e.g. nanomechanical resonators \cite{NatureComm-2011} may be used for
such tests. Various generalizations of Leggett-Garg inequalities are
actively discussed nowadays \cite{Kofler-2013}, as well as their
possible applications to particle physics
\cite{Gangopadhyay:2013aha}. However none of the Leggett-Garg
inequalities are suitable for tests of Bohr's complementarity
principle; it is necessary to find a class of inequalities which
combine the principle of local realism with time dependence. 

Many works \cite{Privitera:1992xc} -- \cite{Donadi:2012nv}
attempt to incorporate time dependence into classic Bell/Wigner
inequalities. Papers \cite{Uchiyama:1996va} -- \cite{Bertlmann:2001ea}
introduce Bell-like inequalities for neutral pseudoscalar mesons
(usually $K$). Two main topics are considered: first, starting
from \cite{Uchiyama:1996va}, the time-independent inequalities are
constructed in terms of flavour, $CP$-violation, or mass/lifetime
eigenstates, and the time dependence arises during the calculation of
the probabilites in the framework of quantum mechanics. The resulting
inequalities contain $CP$-violation parameters $\varepsilon$ and
$\varepsilon'$. There are also attempts \cite{Bertlmann:2001ea} to
include into the inequalities additional correlation functions which
depend on time difference. In yet another approach 
\cite{Privitera:1992xc} -- \cite{Foadi:2000zz}, the authors derive
special time-dependent inequalities based on the principles of
causality and locality, which can be used to test the joint existence
of two characteristics which can not simultaneously be measured. These
inequalities, however, are not general and are only applicable to the
oscillations of neutral mesons. Finally in 
\cite{Bertlmann:2001ea} -- \cite{Donadi:2012nv}, the time-dependent
inequalities similar to \cite{Clauser:1969ny} are introduced, but
there are certain difficulties with violation of these inequalities in
quantum mechanics.

The main property of all variants of Bell-like inequalities is their
conservation if the joint probabilities are non-negative and the
measurements are local. It is possible, however, to introduce
some macroscopic configurations of measurement devices leading to
violation of the inequalities. There are works with criticism of Bell
inequalities per se \cite{khrennikov, griffiths}, stating that the
derivation of the inequalities is not correct itself due to the mutual
unconformity of their constituent values. To address these concerns
it is necessary to formulate clearly our basic assumtions.
The question of the non-locality of measurements is more or less
resolved by Eberhard's theorem \cite{eberhard}.

Unlike \cite{Privitera:1992xc} -- \cite{Foadi:2000zz}, we do not
try to abandon the concept of Bell inequalities, moreover they
naturally re-appear in our analysis. On the other hand, we do not
introduce additional correlation functions which change the structure
of the inequalities, contrary to the approaches stimulated by
\cite{Bertlmann:2001ea}. %We believe that our approach is more
%universal and may be applied to a wider class of correlated systems in
%external fields.

\section{Bell inequalities in Wigner form for a spin-anticorrelated
  fermion pair in an external field}
\label{sec:1}

Following the logic of works \cite{khrennikov} we use Kolmogorov's
approach to probability. 

Let a pseudoscalar particle decay at the moment $t_0$ to a
fermion-antifermion pair. Below we denote the antifermion with the index
``1'' and the fermion with the index ``2''. Let the spin projections
of the fermion and antifermion onto three non-parallel directions $\vec
a$, $\vec b$ and $\vec c$ be simultaneously the elements of 
physical reality. Let us define the spin $1/2$ projection onto any of
the axes 
$\vec n$ as
$$
s_{\vec n}\, =\,\pm\,\frac{1}{2}\,\equiv\, n_{\pm}.
$$
Let the indices $\{\alpha, \beta, \gamma\} = \{+,\, -\}$. Then the
spin projections at the initial moment of time $t=t_0$ onto any
direction are anticorrelated: 
\begin{eqnarray}
\label{pm=mp2}
a_{\pm}^{(1)}(t_0)\, =\, -\, a_{\mp}^{(2)}(t_0).
\end{eqnarray}
Note that in QFT the above condition is automatically satisfied in the
case of a strong or electromagnetic decay with $P$-conservation. It is
easy to construct an example of a Hamiltonian which provides full
anticorrelation at the initial moment of time:
\begin{eqnarray}
\label{Heff_for_PS2ff}
\mathcal{H}^{(PS)}(x)\, =\, g\,\varphi (x)\,\left (\bar f(x)\,\gamma^5\, f(x)\right )_N, 
\end{eqnarray}
where $\varphi (x)$  is a pseudoscalar field and $\bar f(x)$ and $f(x)$ are
fermionic fields.

Let $\Omega$ be the set of elementary outcomes $\omega_i$. For each
of them, the whole set of spin projections is the element of physical reality:
$
\{
a^{(1)}_{\alpha}  b^{(1)}_{\beta}  c^{(1)}_{\gamma}\, 
a^{(2)}_{\alpha'} b^{(2)}_{\beta'} c^{(2)}_{\gamma'}
\}
$.
This set is time-independent. 

At time $t=t_0$ let us define the events 
$
\mathcal{K}_{a^{(1)}_{\alpha} b^{(1)}_{\beta} c^{(1)}_{\gamma}\,
             a^{(2)}_{- \alpha} b^{(2)}_{- \beta} c^{(2)}_{- \gamma}}(t_0)
\subseteq \Omega
$
such that the elements of the physical reality are the sets of
projections of a fermion-antifermion spin pair on three non-parallel
directions, anticorrelated as (\ref{pm=mp2}) 
$
\{
a^{(1)}_{\alpha} b^{(1)}_{\beta} c^{(1)}_{\gamma}\, 
a^{(2)}_{- \alpha} b^{(2)}_{- \beta} c^{(2)}_{- \gamma}
\}
$.
The set of such events by definition forms a $\sigma$-algebra $\mathcal{F}(t_0)$. 
It is possible to introduce a probability measure $w$ on  $(\Omega,\,
\mathcal{F})$ which is real and always non-negative. It is also 
additive ($\sigma$-additive) for non-overlapping events. Then it is
possible to prove the inequality
\begin{eqnarray}
\label{W-B-3}
w\left (a^{(2)}_+, b^{(1)}_+, t_0 \right )\,\le\, 
w\left (c^{(2)}_+, b^{(1)}_+, t_0 \right )\, +\,
w\left (a^{(2)}_+, c^{(1)}_+, t_0 \right ). 
\end{eqnarray}
If one drops the $t_0$  in (\ref{W-B-3}), it becomes identical to
the well-known inequality
\begin{eqnarray}
\label{W-B-2}
w\left (a^{(2)}_+, b^{(1)}_+\right )\,\le\, 
w\left (c^{(2)}_+, b^{(1)}_+\right )\, +\,
w\left (a^{(2)}_+, c^{(1)}_+\right ). 
\end{eqnarray}

To prove (\ref{W-B-3}) and, consequently (\ref{W-B-2}), it is
necessary to consider events 
\begin{eqnarray}
\mathcal{A}(t_0) &=& 
\mathcal{K}_{a^{(1)}_- b^{(1)}_+ c^{(1)}_+\, 
             a^{(2)}_+ b^{(2)}_- c^{(2)}_-}(t_0)\,\cup\,
\mathcal{K}_{a^{(1)}_- b^{(1)}_+ c^{(1)}_-\, 
             a^{(2)}_+ b^{(2)}_- c^{(2)}_+}(t_0),\nonumber \\ 
\mathcal{B}(t_0) &=& 
\mathcal{K}_{a^{(1)}_- b^{(1)}_+ c^{(1)}_-\, 
             a^{(2)}_+ b^{(2)}_- c^{(2)}_+}(t_0)\,\cup\,
\mathcal{K}_{a^{(1)}_+ b^{(1)}_+ c^{(1)}_-\, 
             a^{(2)}_- b^{(2)}_- c^{(2)}_+}(t_0),\nonumber \\
\mathcal{C}(t_0) &=& 
\mathcal{K}_{a^{(1)}_- b^{(1)}_+ c^{(1)}_+\, 
             a^{(2)}_+ b^{(2)}_- c^{(2)}_-}(t_0)\,\cup\,
\mathcal{K}_{a^{(1)}_- b^{(1)}_- c^{(1)}_+\, 
             a^{(2)}_+ b^{(2)}_+ c^{(2)}_-}(t_0),\nonumber 
\end{eqnarray}
belonging to the $\sigma$-algebra $\mathcal{F}(t_0)$. Among all the
indices 
$
\{
a^{(1)}_{\alpha} b^{(1)}_{\beta} c^{(1)}_{\gamma}\, 
a^{(2)}_{- \alpha} b^{(2)}_{- \beta} c^{(2)}_{- \gamma}
\},
$
it is possible to fix not six but only three, which relate to the spin
projection of any fermion to each of three axes. Then 
\begin{eqnarray}
w\left (a_+ ^{(2)}, b_+^{(1)},\, t_0\right ) &=& \sum_{\omega_i \in \mathcal{A}(t_0)}\, 
\left (
w\left (a_+^{(2)}, b_+^{(1)}, c_+^{(2)}, \omega_i \right )\, +\, 
w\left (a_+^{(2)}, b_+^{(1)}, c_-^{(2)}, \omega_i \right )
\right );\nonumber\\
w\left (c_+^{(2)},\, b_+^{(1)}, \, t_0\right ) &=& \sum_{\omega_j \in \mathcal{B}(t_0)}\, 
\left (
w\left (a_+^{(2)}, b_+^{(1)}, c_+^{(2)}, \omega_j \right )\, +\, 
w\left (a_-^{(2)}, b_+^{(1)}, c_+^{(2)}, \omega_j \right )
\right );\nonumber\\
w\left (a_+^{(2)}, c_+^{(1)},\, t_0\right ) &=& \sum_{\omega_k \in \mathcal{C}(t_0)}\, 
\left (
w\left (a_+^{(2)}, b_+^{(1)}, c_-^{(2)}, \omega_k \right )\, +\, 
w\left (a_+^{(2)}, b_-^{(1)}, c_-^{(2)}, \omega_k \right )
\right ).\nonumber
\end{eqnarray}  
The sum $w\left (c_+^{(2)}, b_+^{(1)},\, t_0\right ) + w\left (a_+^{(2)},
  c_+^{(1)},\, t_0\right )$ is defined on the set 
\begin{eqnarray}
\mathcal{B}(t_0)\,\cup\,\mathcal{C}(t_0)\, = \nonumber 
%\end{eqnarray}
%\vspace{-0.6cm}
%\begin{eqnarray}
\left (
  \mathcal{K}_{a^{(1)}_- b^{(1)}_+ c^{(1)}_- a^{(2)}_+ b^{(2)}_- c^{(2)}_+}(t_0) \cup 
  \mathcal{K}_{a^{(1)}_+ b^{(1)}_+ c^{(1)}_- a^{(2)}_- b^{(2)}_- c^{(2)}_+}(t_0)
\right ) \cup \\ \nonumber
\cup \left (
  \mathcal{K}_{a^{(1)}_- b^{(1)}_+ c^{(1)}_+ a^{(2)}_+ b^{(2)}_- c^{(2)}_-}(t_0) \cup
  \mathcal{K}_{a^{(1)}_- b^{(1)}_- c^{(1)}_+ a^{(2)}_+ b^{(2)}_+ c^{(2)}_-}(t_0)
\right ), \nonumber 
\end{eqnarray} 
a subset of which is the event $\mathcal{A}(t_0)$. Then due to the 
non-negativity of probabilities, (\ref{W-B-3}) is proven.

By changing the directions of the axes $\vec a$ and $\vec b$ to their
opposite, as was done in \cite{Nikitin:2009sr}, it is possible to
obtain three more inequalities, analogous to (\ref{W-B-3}):
\begin{eqnarray}
\label{W-B-3times3}
w\left (a^{(2)}_+, b^{(1)}_-, t_0 \right ) & \le & 
w\left (c^{(2)}_+, b^{(1)}_-, t_0 \right )\, +\,
w\left (a^{(2)}_+, c^{(1)}_+, t_0 \right ); \nonumber \\
w\left (a^{(2)}_-, b^{(1)}_+, t_0 \right ) & \le & 
w\left (c^{(2)}_+, b^{(1)}_+, t_0 \right )\, +\,
w\left (a^{(2)}_-, c^{(1)}_+, t_0 \right ); \\
w\left (a^{(2)}_-, b^{(1)}_-, t_0 \right ) & \le & 
w\left (c^{(2)}_+, b^{(1)}_-, t_0 \right )\, +\,
w\left (a^{(2)}_-, c^{(1)}_+, t_0 \right ). \nonumber 
\end{eqnarray}
 
After the time interval $\Delta t = t - t_0$, let the fermion and
antifermion become spatially well separated. Then, under the
assumption of locality, one can write
\begin{eqnarray}
w\left (a^{(2)}_+, b^{(1)}_+, t \right ) 
&=& 
w\left (a^{(2)}_+(t_0) \to a^{(2)}_+ (t) \right )\, 
w\left (b^{(1)}_+(t_0) \to b^{(1)}_+ (t) \right )\,
w\left (a^{(2)}_+, b^{(1)}_+, t_0 \right )\, + \nonumber \\
&+&
w\left (a^{(2)}_-(t_0) \to a^{(2)}_+ (t) \right )\, 
w\left (b^{(1)}_+(t_0) \to b^{(1)}_+ (t) \right )\,
w\left (a^{(2)}_-, b^{(1)}_+, t_0 \right )\, + \nonumber \\
&+&
w\left (a^{(2)}_+(t_0) \to a^{(2)}_+ (t) \right )\, 
w\left (b^{(1)}_-(t_0) \to b^{(1)}_+ (t) \right )\,
w\left (a^{(2)}_+, b^{(1)}_-, t_0 \right )\, + \nonumber \\
&+&
w\left (a^{(2)}_-(t_0) \to a^{(2)}_+ (t) \right )\, 
w\left (b^{(1)}_-(t_0) \to b^{(1)}_+ (t) \right )\,
w\left (a^{(2)}_-, b^{(1)}_-, t_0 \right ).\nonumber 
\end{eqnarray} 
Using the inequalities (\ref{W-B-3}) and (\ref{W-B-3times3}) one can
obtain the following inequality:
\begin{eqnarray}
&& w\left (a^{(2)}_+, b^{(1)}_+, t \right )\,\le \nonumber \\
&\le& 
w\left (a^{(2)}_+(t_0) \to a^{(2)}_+ (t) \right )\, 
w\left (b^{(1)}_+(t_0) \to b^{(1)}_+ (t) \right )\,
\left (
w\left (c^{(2)}_+, b^{(1)}_+, t_0 \right )\, +\,
w\left (a^{(2)}_+, c^{(1)}_+, t_0 \right ) 
\right )\, + \nonumber \\
&+&
w\left (a^{(2)}_-(t_0) \to a^{(2)}_+ (t) \right )\, 
w\left (b^{(1)}_+(t_0) \to b^{(1)}_+ (t) \right )\,
\left (
w\left (c^{(2)}_+, b^{(1)}_+, t_0 \right )\, +\,
w\left (a^{(2)}_-, c^{(1)}_+, t_0 \right )
\right )\, +\nonumber \\
&+&
w\left (a^{(2)}_+(t_0) \to a^{(2)}_+ (t) \right )\, 
w\left (b^{(1)}_-(t_0) \to b^{(1)}_+ (t) \right )\,
\left (
w\left (c^{(2)}_+, b^{(1)}_-, t_0 \right )\, +\,
w\left (a^{(2)}_+, c^{(1)}_+, t_0 \right )
\right )\, + \nonumber \\
&+&
w\left (a^{(2)}_-(t_0) \to a^{(2)}_+ (t) \right )\, 
w\left (b^{(1)}_-(t_0) \to b^{(1)}_+ (t) \right )\,
\left (
w\left (c^{(2)}_+, b^{(1)}_-, t_0 \right )\, +\,
w\left (a^{(2)}_-, c^{(1)}_+, t_0 \right )
\right )\, = \nonumber \\
&=&
w\left (a^{(2)}_+(t_0) \to a^{(2)}_+ (t) \right )\,
\left (
w\left (b^{(1)}_+(t_0) \to b^{(1)}_+ (t) \right )\, +\, 
w\left (b^{(1)}_-(t_0) \to b^{(1)}_+ (t) \right )
\right )\, 
w\left (a^{(2)}_+, c^{(1)}_+, t_0 \right )\, +\nonumber\\
&+&
w\left (a^{(2)}_-(t_0) \to a^{(2)}_+ (t) \right )\,
\left (
w\left (b^{(1)}_+(t_0) \to b^{(1)}_+ (t) \right )\, +\, 
w\left (b^{(1)}_-(t_0) \to b^{(1)}_+ (t) \right )
\right )\, 
w\left (a^{(2)}_-, c^{(1)}_+, t_0 \right )\, +\nonumber\\
&+&
w\left (b^{(1)}_+(t_0) \to b^{(1)}_+ (t) \right )\,
\left (
w\left (a^{(2)}_+(t_0) \to a^{(2)}_+ (t) \right )\, +\, 
w\left (a^{(2)}_-(t_0) \to a^{(2)}_+ (t) \right )
\right )\,
w\left (c^{(2)}_+, b^{(1)}_+, t_0 \right )\, + \nonumber\\ 
&+&
w\left (b^{(1)}_-(t_0) \to b^{(1)}_+ (t) \right )\,
\left (
w\left (a^{(2)}_+(t_0) \to a^{(2)}_+ (t) \right )\, +\, 
w\left (a^{(2)}_-(t_0) \to a^{(2)}_+ (t) \right )
\right )\,
w\left (c^{(2)}_+, b^{(1)}_-, t_0 \right ).\nonumber 
\end{eqnarray}
Thus, due to the existence of an interation, the Bell inequalities in
Wigner form gain time dependence and they can then be written as the
following inequality:
\begin{eqnarray}
\label{W-B-4}
&& w\left (a^{(2)}_+, b^{(1)}_+, t \right )\,\le \\
&\le&
w\left (a^{(2)}_+(t_0) \to a^{(2)}_+ (t) \right )\,
\left (
w\left (b^{(1)}_+(t_0) \to b^{(1)}_+ (t) \right )\, +\, 
w\left (b^{(1)}_-(t_0) \to b^{(1)}_+ (t) \right )
\right )\, 
w\left (a^{(2)}_+, c^{(1)}_+, t_0 \right )\, +\nonumber\\
&+&
w\left (a^{(2)}_-(t_0) \to a^{(2)}_+ (t) \right )\,
\left (
w\left (b^{(1)}_+(t_0) \to b^{(1)}_+ (t) \right )\, +\, 
w\left (b^{(1)}_-(t_0) \to b^{(1)}_+ (t) \right )
\right )\, 
w\left (a^{(2)}_-, c^{(1)}_+, t_0 \right )\, +\nonumber\\
&+&
w\left (b^{(1)}_+(t_0) \to b^{(1)}_+ (t) \right )\,
\left (
w\left (a^{(2)}_+(t_0) \to a^{(2)}_+ (t) \right )\, +\, 
w\left (a^{(2)}_-(t_0) \to a^{(2)}_+ (t) \right )
\right )\,
w\left (c^{(2)}_+, b^{(1)}_+, t_0 \right )\, + \nonumber\\ 
&+&
w\left (b^{(1)}_-(t_0) \to b^{(1)}_+ (t) \right )\,
\left (
w\left (a^{(2)}_+(t_0) \to a^{(2)}_+ (t) \right )\, +\, 
w\left (a^{(2)}_-(t_0) \to a^{(2)}_+ (t) \right )
\right )\,
w\left (c^{(2)}_+, b^{(1)}_-, t_0 \right ).\nonumber 
\end{eqnarray}
We would like to emphasise that  (\ref{W-B-4}) is proved on the set of
elementary outcomes $\Omega$, which is time-independent.

In the absence of any interactions,  
$w\left (a^{(2)}_-(t_0) \to a^{(2)}_+ (t) \right ) =  w\left
  (b^{(1)}_-(t_0) \to b^{(1)}_+ (t) \right ) = 0$, while 
$w\left (a^{(2)}_+(t_0) \to a^{(2)}_+ (t) \right ) = w\left
  (b^{(1)}_+(t_0) \to b^{(1)}_+ (t) \right ) = 1$. Hence
(\ref{W-B-4}) devolves to (\ref{W-B-3}), as it should from the
physical point of view. The inequality (\ref{W-B-3}) is in turn
equivalent to the time-independent inequality (\ref{W-B-2}). 

The time-dependent inequality (\ref{W-B-4}) is the main result of this 
paper. In the next sections we will show how it can be applied to some
real experimental situations. We will demonstrate some advantages of
the time-dependent inequality  (\ref{W-B-4}) over the static
inequality (\ref{W-B-2}).

%============================================================================================

\section{NRQM: anticorrelated spins in an external magnetic field}
\label{sec:2}

Let us consider the following example.

Let a pseudoscalar particle at rest decay to a positron (index ``1'')
and an electron (index ``2''). It is easy to show (see
e.g. \cite{Nikitin:2009sr}), that the $e^+ e^-$ pair is in the state
with zero momentum and spin. Let us suppose that at the 
time $t_0 = 0$, the spins of the electron and positron are fully
anticorrelated along the axis $z$. Then the spin wave function of the
$e^+ e^-$ pair at $t=t_0=0$ may be written as:

\begin{eqnarray}
\label{correlation-t=0}
\ket{\Psi (t=0)}\, =\,\frac{1}{\sqrt{2}}\, 
\left [
\left (
\begin{array}{c}
     1 \\
     0  
\end{array}
\right )^{(2)}
\left (
\begin{array}{c}
     0 \\
     1  
\end{array}
\right )^{(1)}\, -\,
\left (
\begin{array}{c}
     0 \\
     1  
\end{array}
\right )^{(2)}
\left (
\begin{array}{c}
     1 \\
     0  
\end{array}
\right )^{(1)}
\right ].
\end{eqnarray}
Let us suppose that the electron-positron system is embedded in a
constant and homogeneous magnetic field $\vec{\mathcal{H}}$, 
which is directed along the axis $y$. Let us measure the spin
projections in the $(x,z)$ plane on three non-parallel axes $\vec a$,
$\vec b$, and $\vec c$. Also let us require the particles to propagate
along the $y$ axis to prevent rotation of these charged particles in the
magnetic field. The spins of the fermions will begin
to precess around the $y$ axis. Let us consider this precession for
particles which have a definite spin projection onto the direction 
$\vec a$. In a spherical coordinate system at time $t=t_0=0$, the
corresponding electron and positron wave functions are:
%\begin{equation}
\[
\ket{\frac{1}{2},\, a_+^{(i)}} = \left (
\begin{array}{c}
     \cos  \theta_a/2\\
     \sin  \theta_a/2
\end{array}
\right )\qquad \textrm{and}\qquad
\ket{\frac{1}{2},\, a_-^{(i)}} = \left (
\begin{array}{r}
     -\,\sin  \theta_a/2\\
     \cos  \theta_a/2
\end{array}
\right ), %\nonumber
\]
%\end{equation}
where $i = \{1,\, 2\}$. The wave function of the electron at an
arbitrary time is: 
\begin{eqnarray}
\label{wf-t-e-}
\ket{\psi^{(2)}_{a_+}(t)} = \left (
\begin{array}{c}
     \cos  \left (\omega t + \theta_a/2 \right )\\
     \sin   \left (\omega t + \theta_a/2 \right )
\end{array}
\right )^{(2)}\qquad \textrm{and}\qquad
\ket{\psi^{(2)}_{a_-}(t)} =
\left (
\begin{array}{r}
     -\,\sin  \left (\omega t + \theta_a/2 \right )\\
        \cos  \left (\omega t + \theta_a/2 \right )
\end{array}
\right )^{(2)},
\end{eqnarray}
and for the positron wave function: 
\begin{eqnarray}
\label{wf-t-e+}
\ket{\psi^{(1)}_{a_+}(t)} = \left (
\begin{array}{r}
     \cos  \left (\omega t - \theta_a/2 \right )\\
     -\,\sin  \left (\omega t - \theta_a/2 \right )
\end{array}
\right )^{(1)}\qquad \textrm{and}\qquad
\ket{\psi^{(1)}_{a_-}(t)} =
\left (
\begin{array}{c}
        \sin  \left (\omega t - \theta_a/2 \right )\\
        \cos  \left (\omega t - \theta_a/2 \right )
\end{array}
\right )^{(1)},
\end{eqnarray}
where $\displaystyle\omega = \frac{e \mathcal{H}}{2 m_e c}$ is the
Larmor precession frequency of the electron.

Taking into account the initial condition (\ref{correlation-t=0}) and the
electron and positron wave functions (\ref{wf-t-e-}), (\ref{wf-t-e+})
in the magnetic field, and assuming $\theta_z=0$, we obtain the wave
function of the $e^+e^-$ pair for an arbitrary time $t$:
\begin{eqnarray}
\ket{\Psi (t)}\, =\,\frac{1}{\sqrt{2}}\, 
\left  [
\left (
\begin{array}{c}
     \cos (\omega t) \\
     \sin (\omega t)  
\end{array}
\right )^{(2)}
\left (
\begin{array}{c}
     \sin (\omega t) \\
     \cos (\omega t)  
\end{array}
\right )^{(1)}\, 
-\, 
\left (
\begin{array}{r}
     -\,\sin (\omega t) \\
        \cos (\omega t)  
\end{array}
\right )^{(2)}
\left (
\begin{array}{r}
        \cos (\omega t) \\
     -\,\sin (\omega t)  
\end{array}
\right )^{(1)}\, 
\right ]. \nonumber
\end{eqnarray}
Now let us compute all the probabilities that entering the inequality (\ref{W-B-4}):
\begin{eqnarray}
\label{probH-1}
&& w(a^{(2)}_+,\, b^{(1)}_+,\, t)\, =\,
\left |\bra{\frac{1}{2},\, a^{(2)}_+}\bracket{\frac{1}{2},\, b^{(1)}_+}{\Psi (t)}\right |^2\, =\, 
\frac{1}{2}\,\sin^2\left (\frac{\theta_{ba}}{2}\, +\, 2 \omega t \right),\nonumber \\ 
&& w(a^{(2)}_-,\, c^{(1)}_+,\, t)\, =\,
\left |\bra{\frac{1}{2},\, a^{(2)}_-}\bracket{\frac{1}{2},\, c^{(1)}_+}{\Psi (t)}\right |^2\, =\, 
\frac{1}{2}\,\cos^2\left (\frac{\theta_{ca}}{2}\, +\, 2 \omega t \right), \\
&& w(c^{(2)}_+,\, b^{(1)}_-,\, t)\, =\,
\left |\bra{\frac{1}{2},\, c^{(2)}_+}\bracket{\frac{1}{2},\, b^{(1)}_-}{\Psi (t)}\right |^2\, =\, 
\frac{1}{2}\,\cos^2\left (\frac{\theta_{cb}}{2}\, -\, 2 \omega t \right). \nonumber 
\end{eqnarray}
Then:
\begin{eqnarray}
\label{probH-2}
&&w \left ( a_+^{(2)}(t_0=0) \to a_+^{(2)}(t) \right ) = \left | \bracket{\psi^{(2)}_{a+}(t)}{\frac{1}{2},\, a_+^{(2)}}\right |^2 = \cos^2 \left ( \omega t \right ), \nonumber \\
&&w \left ( a_-^{(2)}(t_0=0) \to a_+^{(2)}(t) \right ) = \left | \bracket{\psi^{(2)}_{a-}(t)}{\frac{1}{2},\, a_+^{(2)}}\right |^2 = \sin^2 \left ( \omega t \right ),\\
&&w \left ( b_+^{(1)}(t_0=0) \to b_+^{(1)}(t) \right ) = \left | \bracket{\psi^{(1)}_{b+}(t)}{\frac{1}{2},\, b_+^{(1)}}\right |^2 = \cos^2 \left ( \omega t \right ), \nonumber \\
&&w \left ( b_-^{(1)}(t_0=0) \to b_+^{(1)}(t) \right ) = \left | \bracket{\psi^{(1)}_{b-}(t)}{\frac{1}{2},\, b_+^{(1)}}\right |^2 = \sin^2 \left ( \omega t \right ). \nonumber
\end{eqnarray}
Substituting (\ref{probH-1}) and (\ref{probH-2}) into (\ref{W-B-4}),
we obtain the Bell inequality in Wigner form for our model:
\begin{eqnarray}
\label{main_en}
\sin^2\left (\frac{\theta_{ba}}{2}\, +\, 2 \omega t\right )\,\le\, 
2 \sin^2 \left ( \omega t \right )\, +\, 
  \cos \left (2 \omega t \right )\, 
     \left [
              \sin^2\left (\frac{\theta_{ca}}{2}\right )\, +\, 
              \sin^2\left (\frac{\theta_{bc}}{2}\right )
     \right ].
\end{eqnarray} 
Note that if at the initial time $t_0=0$ the spins of the 
electron and positron are correlated along any arbitrary direction
$\vec n$, instead of the $z$ axis, then the derivation of
(\ref{main_en}) becomes more complex, but ultimately its structure does
not change. This fact might not be obvious initially, because of
the unique direction in the example which is determined by the
direction of the magnetic field $\vec{\mathcal{H}}$.

Let us estimate the level of possible violation of inequality
(\ref{main_en}) assuming one is free to choose the directions of the
spin projections and the magnetic field strength. Before we consider
the general case, let us examine two important particular
cases.

The first case: let $\omega t = n \pi$, where $n = 0,\, 1,\, 2\,
\ldots$. Then (\ref{main_en}) becomes the classic time-independent
inequality
\begin{eqnarray}
\sin^2\left (\frac{\theta_{ba}}{2} \right)\,\le\, 
\sin^2\left (\frac{\theta_{ca}}{2}\right )\, +\, 
\sin^2\left (\frac{\theta_{bc}}{2}\right ),\nonumber
\end{eqnarray} 
for which the maximum of violation is reached if $\displaystyle\theta_{bc} =
\theta_{ca} = \frac{\pi}{3}$ and $\displaystyle\theta_{ba} =
\theta_{bc} + \theta_{ca} = \frac{2 \pi}{3}$.  With these values we
end up with a false inequality $\displaystyle\frac{3}{4} \le
\frac{1}{2}$.

The second case appears if $\displaystyle\omega t = \frac{\pi}{2} +
\pi n$. Then (\ref{main_en}) turns into inequality
\begin{eqnarray}
\sin^2\left (\frac{\theta_{ba}}{2} \right)\, +\, 
\sin^2\left (\frac{\theta_{ca}}{2}\right )\, +\, 
\sin^2\left (\frac{\theta_{bc}}{2}\right )\,\le\, 2, \nonumber
\end{eqnarray} 
for which the maximum of violation $\displaystyle\left ( \frac{9}{4} \le
  2\,\,\textrm{or}\,\,\frac{1}{4} \le 0 \right )$ is reached  with
$\displaystyle\theta_{bc} = \theta_{ca} = \frac{2
  \pi}{3}$ and $\displaystyle\theta_{ba} = \theta_{bc} + \theta_{ca} =
\frac{4 \pi}{3}$, i.e. with the most symmetric configuration of axes
$\vec a$, $\vec b$, and $\vec c$ in the plane $(x,\, z)$. A similar
inequality may be obtained for free correlated particles, see
e.g. \cite{Nikitin:2009sr}.

It is obvious that is both the above cases, the inequality
(\ref{main_en}) does not have any advantage over the classic
inequality (\ref{W-B-2}). Is it possible to violate (\ref{main_en})
more than (\ref{W-B-2}) using the magnetic field?

Let us choose the angle $\theta$ so that $\displaystyle\cos (2
\theta) = \frac{1}{4}$ (i.e., $\theta \approx
37,8^o$), and let the angles between the spin projections and the
magnetic field direction satisfy the following: $\displaystyle\theta_{bc} =
\theta_{ca} = \theta$, $\displaystyle\theta_{ba} = \theta_{bc} +
\theta_{ca} = 2 \theta$, and $\displaystyle\omega t = \frac{\theta}{2} +
\pi n$. In this case the maximal violation of (\ref{main_en})
becomes $\displaystyle\frac{9}{16} \le 0$. This violation exceeds the 
violation of (\ref{W-B-2}) by more than a factor of two. 

This result is a consequence of the new degree of freedom provided by
the magnetic field.

Note, that every real measurement device has a finite time resolution
$\Delta t$. To take this into account, let us average (\ref{main_en})
using operator
%\begin{equation}
%\nonumber
\[
	I_\delta[f(\tau)]\, =\,\frac{1}{2\delta}\,\int\limits_{T-\delta}^{T+\delta} \, f(\tau)\, d\tau ,
\]
%\end{equation}
where $2\delta = \omega \Delta t$ is the dimensionless characteristic
resolution of the macro-device,  $\tau = \omega t$ is a dimensionless
time parameter. The $T$ is a dimensionless ``common time.'' 
The operator $I_\delta[f(\tau)]$ affects the summands of
(\ref{main_en}) as follows: 
\newcommand{\sinc}{\mathop{\mathrm{sinc}}}
\begin{eqnarray*}
	&&I_\delta\left[\sin^2\left (\frac{\theta_{ba}}{2}\, +\, 2 \tau \right)\right] = 
		\frac{1}{2}\Big(1-\sinc(4\delta)\Big) \;+\; \sinc(4\delta)  \,
				\sin^2\left (\frac{\theta_{ba}}{2}\, +\, 2 T \right); \\
	&&I_\delta\left[\sin^2\left ( \tau \right )\right] = 
		\frac{1}{2}\Big(1-\sinc(2\delta)\Big) \;+\; \sinc(2\delta)  \,
				\sin^2\left( T \right); \\
	&&I_\delta\left[\cos\left ( 2\tau \right )\right] = \sinc(2\delta)  \,
				\cos\left(2 T \right),
\end{eqnarray*}
where the cardinal sine $\sinc (x) = \sin (x)/x$. Substituting these
expressions into (\ref{main_en}), we obtain an inequality which
accounts for the finite time resolution of a macro-device:
\begin{eqnarray}
	\label{discr1}
	\sinc(2\delta) \;  \left[ \cos(2\delta)\left( \sin^2\left
          (\frac{\theta_{ba}}{2}\, +\, 2 T \right) -\frac{1}{2}
      \right)\;+\; 1\,-\, 2\sin^2\left( T
      \right)\,-\,\right. \nonumber \\
	 \left. \cos\left(2 T \right) 
				\left( \sin^2\left (\frac{\theta_{ca}}{2}\right )\, +\, 
				              \sin^2\left (\frac{\theta_{bc}}{2}\right
                              )\right)\right] \;\leq\; \frac{1}{2}.
\end{eqnarray}

For an ideal $(\delta\rightarrow 0)$ macro-device this inequality 
becomes inequality (\ref{main_en}). If the time resolution of the device
substantially exceeds the precession period ($\delta>>1$), then the
inequality (\ref{discr1}) turns into the trivial expression $0\leq
1/2$. 

\begin{figure}[h]
\label{fig:1}
\begin{center}
%\begin{tabular}{c}
%\mbox{\epsfig{file=func.pdf,width=14.cm}}
\includegraphics{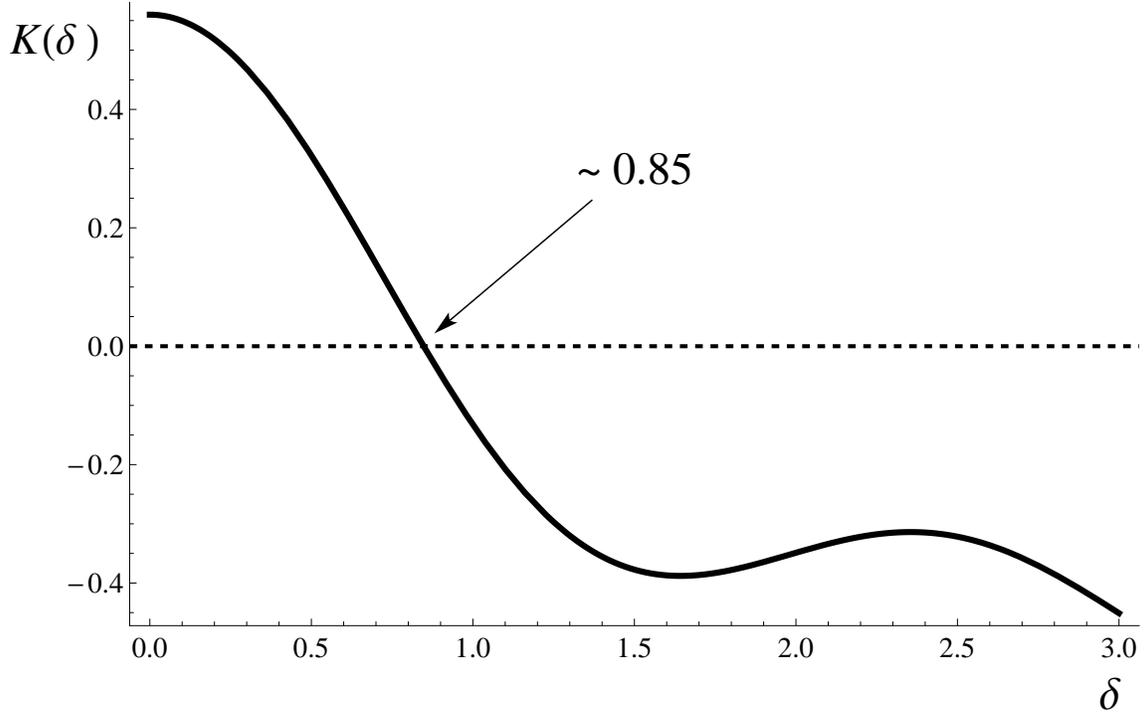}
%\end{tabular}
\caption{Violation rate of the inequality
    \protect\ref{discr1} versus the time resolution.}
\end{center}
\end{figure}

Let us apply \ref{discr1} to the case of maximum violation of
the inequality (\ref{main_en}) when $\displaystyle\cos (2 \theta) =
\frac{1}{4}$.
Then  \ref{discr1} goes into the inequality 
\begin{equation}
	\mathop{\mathit{K}}(\delta) = \sinc(2\delta)\Big(0.44\,
    \cos(2\delta) \,+\, 0.62 \Big) - \frac{1}{2} \;\leq\; 0.
\end{equation}
This inequality depends only on the dimensionless time resolution of
the macro-device. The function $\mathop{\mathit{K}}(\delta)$
is presented in Fig. \ref{fig:1}. One can see that to
experimentally resolve the violation of (\ref{main_en}), it is
necessary to have $\delta\lesssim 0.85$. Then the time resolution of
the macro-device may be estimated as
$$
\Delta t \,\lesssim\, \frac{1.7}{\omega}.
$$
Considering a magnetic field $\mathcal{H} = 1$ T, one can get for an
electron $\Delta t \lesssim 10^{-13}$ s, and for a proton $\Delta t
\lesssim 10^{-7}$ s. These time thresholds might be increased by
decreasing the magnetic field strength. 

%============================================================================================

\section{Time-dependent Bell inequalities in Wigner form for
  oscillations of neutral pseudoscalar mesons}
\label{sec:3}
{
Despite the fact that in Section \ref{sec:1} we
used spin projections of fermions, the inequality (\ref{W-B-4}) is
correct for any three dichotomic observables. The only requirement for
tests of Bohr's principle of complementarity is for corresponding
operators not to commute with each other.

To demonstrate the distinction between the inequalities (\ref{W-B-4})
and (\ref{W-B-2}), let us consider as an example the oscillations of
neutral $B$-mesons, where the static inequalities 
(\ref{W-B-2}) are never violated, while the dynamical inequalities 
(\ref{W-B-4}) are violated almost always.

The main idea of static Bell inequalities in Wigner form was
introduced in 
\cite{Privitera:1992xc,Uchiyama:1996va,Bramon:1998nz} and was
further developed in
\cite{Bertlmann:2001sk,Bertlmann:2001ea,Bramon:2005mg}. 
Its gist is to select three distinct ``directions'' in the system of
neutral $B$-mesons, whose operators do not commute with each
other. The first ``direction'' is the $B$-meson flavour, i.e. the
projection onto the states $\ket{B} = \ket{\bar b q}$ and $\ket{\bar B}
= \ket{b \bar q}$, where $q=\{ d,\, s\}$.  
Let us define the operators of charge ($\hat C$) and space ($\hat P$)
conjugation for these states as
\begin{eqnarray}
\hat C \hat P\,\ket{B}\, =\, e^{i\alpha} \ket{\bar B}\quad\textrm{and}\quad\hat C \hat P\,\ket{\bar B}\, =\, e^{-i \alpha} \ket{B}, \nonumber
\end{eqnarray}
where $\alpha$ is a non-physical arbitrary phase of the
$CP$-violation. This phase should not appear in any experimentally
testable inequalities.

The second ``direction'' is the states with a
definite value of $CP$, i.e. the states
\begin{eqnarray}
\ket{B_1} = \frac{1}{\sqrt{2}}\,\left ( \ket{B} - e^{i\alpha} \ket{\bar B}\right ), \quad 
\ket{B_2} = \frac{1}{\sqrt{2}}\,\left ( \ket{B} + e^{i\alpha} \ket{\bar B}\right ), \nonumber
\end{eqnarray}
with negative and positive $CP$ accordingly. 

The third ``direction'' is defined by the mass and lifetime eigenstates:
\begin{eqnarray}
\ket{B_L} = p\ket{B} + q \ket{\bar B}\quad\textrm{and}\quad\ket{B_H} = p\ket{B} - q \ket{\bar B}. \nonumber
\end{eqnarray}
The latter two states are the eigenvectors of a non-Hermitan
Hamiltonian, with eigenvalues of $E_L = m_L - i \Gamma_L/2$ and $E_H =
m_H - i \Gamma_H/2$ respectively (we use $\hbar = c
=1$). These two states are not orthogonal to each other. The complex
coefficients $p$ and $q$ are normalized in the standard way:
\begin{eqnarray}
|p|^2 + |q|^2 = |\tilde p|^2 + |q|^2 =1, \nonumber
\end{eqnarray}
where $\tilde p = p e^{i\alpha}$. The decay $\Upsilon (4S) \to B \bar B$
defines the wave function of the $B \bar B$-system in the flavour space at
$t = t_0$:
\begin{eqnarray}
\label{correlationBbarB-t=0}
\ket{\Psi (t_0)} = \frac{1}{\sqrt{2}}
\left (\ket{B}^{(2)} \ket{\bar B}^{(1)}\, -\,\ket{\bar B}^{(2)} \ket{B}^{(1)}\right ),
\end{eqnarray}
which is identical to the wave function (\ref{correlation-t=0})
in spin space. In an experiment one can distinguish between the
$B$-mesons by their direction in the detector (this procedure is
detailed in \cite{Uchiyama:1996va,Bertlmann:2001sk,Bertlmann:2001ea}).

Let us briefly overview the case of the time-independent inequalities. Here
we follow the logic of \cite{Bertlmann:2001ea}.
Let us correspond: $a_+ \to B_1$, $a_- \to B_2$, $b_+ \to
\bar B$, $b_- \to B$, $c_+ \to B_H$ and $c_- \to B_L$. 
Then the classic inequality (\ref{W-B-2}) becomes:
\begin{eqnarray}
\label{W-B-2-BbarB-1}
w(B_1^{(2)},\, \bar B^{(1)},\, t_0)\,\le\, w(B_1^{(2)},\, B_H^{(1)},\, t_0) + w(B_H^{(2)},\, \bar B^{(1)},\, t_0).
\end{eqnarray} 
Using the initial condition 
(\ref{correlationBbarB-t=0}) one finds that:
\begin{eqnarray}
\label{w-BbarB-I}
&& w(B_1^{(2)},\, \bar B^{(1)},\, t_0)\,=\,  \left |\bra{B_1^{(2)}}\bracket{\bar B^{(1)}}{\Psi (t_0)}\right |^2\, =\,
\frac{1}{4}\, =\, \frac{1}{4}\,\left ( |\tilde p|^2 + |q|^2 \right );\nonumber \\
&& w(B_1^{(2)},\, B^{(1)},\, t_0)\,=\,  \left |\bra{B_1^{(2)}}\bracket{B^{(1)}}{\Psi (t_0)}\right |^2\, =\,
\frac{1}{4}\, =\, \frac{1}{4}\,\left ( |\tilde p|^2 + |q|^2 \right );\nonumber \\
&& w(B_1^{(2)},\, B_H^{(1)},\, t_0)\,=\, \left |\bra{B_1^{(2)}}\bracket{\bar B^{(1)}_H}{\Psi (t_0)}\right |^2\, =\,
\frac{1}{4}\, \left |\tilde p - q\right |^2; \\
&& w(B_2^{(2)},\, B_H^{(1)},\, t_0)\,=\, \left |\bra{B_2^{(2)}}\bracket{\bar B^{(1)}_H}{\Psi (t_0)}\right |^2\, =\,
\frac{1}{4}\, \left | \tilde p + q\right |^2; \nonumber \\
&& w(B_H^{(2)},\, \bar B^{(1)},\, t_0)\,=\,\left |\bra{B_H^{(2)}}\bracket{\bar B^{(1)}}{\Psi (t_0)}\right |^2\, =\,
\frac{1}{2}\, \left | \tilde p \right |^2, \nonumber \\
&& w(B_H^{(2)},\, B^{(1)},\, t_0)\,=\,\left |\bra{B_H^{(2)}}\bracket{B^{(1)}}{\Psi (t_0)}\right |^2\, =\,
\frac{1}{2}\, \left | q \right |^2. \nonumber
\end{eqnarray}
Substituting the probabilities (\ref{w-BbarB-I}) into 
(\ref{W-B-2-BbarB-1}) leads to:
\begin{eqnarray}
\label{W-B-2-pq-1}
|q|^2 - |\tilde p|^2 \le \left | \tilde p - q\right |^2,
\end{eqnarray}
which is identical to the inequality (16) from 
\cite{Uchiyama:1996va} for the oscillations of neutral $K$-mesons if
one redefines $p$ and $q$ through the $CP$-violation parameter
$\varepsilon$. As there is not direct dependence between the
projections of the states onto various directions, one can suppose that
$b_+ \to B$ and $b_- \to \bar B$. Then (\ref{W-B-2}) becomes 
\begin{eqnarray}
\label{W-B-2-BbarB-2}
w(B_1^{(2)},\, B^{(1)},\, t_0)\,\le\, w(B_1^{(2)},\, B_H^{(1)},\, t_0) + w(B_H^{(2)},\, B^{(1)},\, t_0),
\end{eqnarray} 
and consequently, taking into account  (\ref{w-BbarB-I}):
\begin{eqnarray}
\label{W-B-2-pq-2}
|\tilde p|^2 - |q|^2 \le \left | \tilde p - q\right |^2.
\end{eqnarray}
The inequalities (\ref{W-B-2-pq-1}) and (\ref{W-B-2-pq-2}) may be
uniformly written as:
\begin{eqnarray}
\label{W-B-2-pq-3}
\left | |\tilde p|^2 - |q|^2 \right |\,\le\, \left | \tilde p - q\right |^2.
\end{eqnarray}
Relation (\ref{W-B-2-pq-3})  is never violated, as follows from the
triangle inequality for complex numbers. The equality is reached when
$|\tilde p| = |q|$ or when $|p| = |q|$. All possible re-definitions of
``directions'' and projections of the $B$-meson states upon them lead
to inequality (\ref{W-B-2-pq-3}). In
\cite{Bertlmann:2001sk,Bertlmann:2001ea} it is supposed that the
inequality similar to (\ref{W-B-2-pq-1}) may be violated by a special
choice of the non-physical phase $\alpha$. But it is obvious that the
inequalities (\ref{W-B-2-pq-1}), (\ref{W-B-2-pq-2}), and
(\ref{W-B-2-pq-3}) do not depend on $\alpha$, as should be the case
for any experimentally testable theory. Hence the static inequalities
(\ref{W-B-2}), written in terms of the oscillation parameters of neutral
$B$-mesons (\ref{W-B-2-BbarB-1}), (\ref{W-B-2-BbarB-2}), are never
violated, and do not allow testing of the complementarity principle.

Let us now consider the time-dependent inequalities (\ref{W-B-4}),
written for a system of neutral $B$-mesons. Note that for the
derivation of (\ref{W-B-4}) the probability normalisation was never
used. Hence (\ref{W-B-4}) is valid for decays in which the 
normalisation of the state vectors depends on time. Again set $a_+ \to
B_1$, $a_- \to B_2$, $b_+ \to \bar B$, $b_- \to B$,
$c_+ \to B_H$ and $c_- \to B_L$. 
The time evolution of the states $\ket{B_L}$ and $\ket{B_H}$ is
trivial: 
\begin{eqnarray}
\ket{B_L (t)} =  e^{- i m_L t - \Gamma_L t/2} \ket{B_L}, \qquad \ket{B_H (t)} =  e^{- i m_H t - \Gamma_H t / 2} \ket{B_H}. \nonumber
\end{eqnarray}
This defines the evolution of the states $\ket{B (t)}$ and $\ket{\bar B
  (t)}$ as follows:
\begin{eqnarray}
\left \{
\begin{array}{l}
\displaystyle \ket{B(t)} = g_+(t) \ket{B}\, -\,\frac{q}{p} g_-(t) \ket{\bar B} \\
\displaystyle \ket{\bar B (t)} = - \frac{p}{q} g_-(t) \ket{B} + g_+(t) \ket{\bar B}
\end{array}
\right . . \nonumber
\end{eqnarray}
Now one can derive the evolution of the state $\ket{B_1(t)}$ as:
\begin{eqnarray}
\ket{B_1(t)} = \frac{1}{\sqrt{2}}
\left (
\left ( 
g_+(t) + \frac{p}{q}\, e^{i \alpha} g_-(t)
\right ) \ket{B}\, -\,
\left ( 
g_+(t)\,  e^{i \alpha} + \frac{q}{p}\, g_-(t)
\right ) \ket{\bar B}
\right ),
\nonumber
\end{eqnarray}
where $\displaystyle g_{\pm}(t) = \frac{1}{2}\,\left ( e^{-i E_H t}
  \pm e^{- i E_L t}\right )$. Functions $g_{\pm}(t)$ satisfy the 
conditions: 
\begin{eqnarray}
&& \left | g_{\pm}(t) \right |^2 = \frac{e^{-\Gamma t}}{2}\,
\left (
\mathrm{ch} \left ( \frac{\Delta\Gamma\, t}{2}\right ) \pm \cos \left
  ( \Delta m t\right ) \right ) \mathrm{ and} \nonumber \\
&& g_+ (t) g_-^*(t) = \frac{e^{-\Gamma t}}{2}\,
\left (
-\mathrm{sh} \left ( \frac{\Delta\Gamma\, t}{2}\right ) +  i\sin \left
  ( \Delta m t\right ) 
\right ),  \nonumber 
\end{eqnarray}
where $\Gamma = (\Gamma_H + \Gamma_L)/2$, $\Delta \Gamma = \Gamma_H -
\Gamma_L$, and $\Delta m = m_H - m_L$. Taking into account initial
condition (\ref{correlationBbarB-t=0}), it is possible to write for the
wave function of the $B\bar B$-pair at an arbitrary moment of time:
\begin{eqnarray}
\label{correlationBbarB-t}
\ket{\Psi (t)} =  e^{-i \left ( m_H + m_L\right )\, t}\, e^{- \Gamma t}\,\ket{\Psi (t_0)}.
\end{eqnarray}
For the neutral $B$-mesons, $\displaystyle \left (\frac{q}{p} \right
)^2 = e^{2 i \alpha}$ \cite{Beringer:1900zz}. 
The experimental value for this ratio is \cite{Beringer:1900zz}:
$$
\left |\frac{q}{p} \right |\, =\, 1.0017 \pm 0.0017.
$$
Considering the normalisation at $t=t_0$, we can let $|p|^2 
\approx |q|^2 \approx 1/2$. Subsequent calculations will be performed
for the case of $q/p = e^{i \alpha}$. Then:
\begin{eqnarray}
\label{w-BbarB-II}
&& w(B_1^{(2)}(0) \to B_1^{(2)}(t)) =  \left |\bracket{B_1(t)}{B_1} \right |^2= e^{- \Gamma\, t}\, e^{- \Delta \Gamma\, t/2}; \nonumber \\
&& w(B_2^{(2)}(0) \to B_1^{(2)}(t)) =  \left |\bracket{B_1(t)}{B_2} \right |^2= 0; \nonumber \\
&& w(\bar B^{(1)}(0) \to \bar B^{(1)}(t)) =  \left |\bracket{\bar B (t)}{\bar B} \right |^2 =  |g_+ (t)|^2; \\
&& w(B^{(1)}(0) \to \bar B^{(1)}(t)) =  \left |\bracket{\bar B (t)}{B} \right |^2 = |g_-(t)|^2; \nonumber \\
&&w(B_1^{(2)},\, \bar B^{(1)},\, t) = \left |\bra{B_1^{(2)}}\bracket{\bar B^{(1)}}{\Psi (t)}\right |^2\, =\,\frac{1}{4}\, e^{-2 \Gamma\, t}. \nonumber
\end{eqnarray}
The substitution of (\ref{w-BbarB-I}) and (\ref{w-BbarB-II}) into
(\ref{W-B-4}) results in the following time-dependent inequality:
\begin{eqnarray}
\label{W-B-2-gammaT}
1\,\le\, e^{-\,\Delta \Gamma\, t}.
\end{eqnarray}
If $\Delta \Gamma = \Gamma_H - \Gamma_L> 0$, then the inequality
(\ref{W-B-2-gammaT}) is violated for any $t  > 0$. Note that
(\ref{W-B-2-gammaT}) does not contradict the static inequality
(\ref{W-B-2-pq-3}), because with $q/p = e^{i\alpha}$ the latter also
reverts to equality.

While the static Bell inequality in Wigner form is never violated in
the example, regardless of the choices of $p$ and $q$, the
dynamical inequality is always violated for $q/p = e^{i \alpha}$, if
$\Delta \Gamma = \Gamma_H - \Gamma_L> 0$.
The latter condition is proven experimentally for the neutral
$B_{d,s}$-mesons \cite{Beringer:1900zz}.
}

%============================================================================================

\section{QFT: Bell inequalities at the leading order}
\label{sec:4}

In the framework of NRQM, the violation of the time-dependent Bell
inequalities in Wigner form (\ref{W-B-4}), like the violation
of the static inequalities (\ref{W-B-2}), is determined by its
non-locality. That is, this violation in NRQM means one of the following:

\begin{itemize}
\item[\textbf{a)}] NRQM is non-local, but the spin projections onto
  non-parallel directions may simultaneously be the elements of the
  physical reality;
\item[\textbf{b)}] NRQM is non-local, but Bohr's complementarity
  principle is valid, i.e. the spin projections depends on the
  configuration of a macro-device. 

\end{itemize}
Hence, in order to test the complementarity principle alone, it is
necessary to exclude non-locality. This is possible if one
considers (\ref{W-B-4}) in the framework of QFT, which is local by
construction. In QFT the experimental proof of a violation of
(\ref{W-B-4}) will mean the validity of the complementarity principle.

Standard methods in QFT only allow calculation of the time-dependent
decay probability $W(t) = \partial w(t)/\partial t$ with perturbative
theory. As an example let us consider a decay of a free neutral
pseudoscalar particle $P$ to a fermion-antifermion pair $f ^+ f^-$. Let
antifermion $f^+$ to have the index ``1'', and fermion $f^-$ to
have the index ``2''. The Hamiltonian of the decay is defined by
expression (\ref{Heff_for_PS2ff}),  where $f(x)$ and $\bar f(x)$ are 
fermionic fields, and $\varphi(x)$ is a pseudoscalar field. Let the
pseudoscalar particle have mass $M$, and let the fermionic
masses to be negligible (this assumption will not affect the final
result but will simplify the calculations). The decay width $\Gamma_0$
when $t_0 \to -\infty$ or $t \to +\infty$ is equal to $\displaystyle\Gamma_0 =
\frac{g^2 M}{8 \pi}$. However the probabilities in (\ref{W-B-4}) are
defined for finite times $t$ and $t_0$. Let us calculate them.

Using the technique of calculations in QFT for finite times
\cite{Bernardini:1993qj,Nomoto:1991vk,Joichi:1997xn}, one obtains the
following expression for the probability of the decay $P \to f^+ f^-$:
\begin{eqnarray}
\label{w1a+b+t}
W^{(1)}(a^{(2)}_+, b^{(1)}_+, \tau)\, =\,\frac{\Gamma_0}{2}\, 
\left (
1\, +\, \frac{\textrm{si} (M\,\tau)}{\pi}\, +\, 
        \frac{\sin (M\,\tau)}{\pi (M\,\tau)^2}\, +\, 
        \frac{\cos (M\,\tau t)}{\pi (M\,\tau t)}
\right )\,\times\,\sin^2 \frac{\theta_{ab}}{2},  
\end{eqnarray}
where $\tau $ is the time of measurement, and $\textrm{si}(x)$ is the 
integral sine, defined as 
$$
\textrm{si}(x) = - \,\int\limits_x^{+\infty} \frac{\sin \zeta}{\zeta}\, d\zeta .
$$

For $\tau \to +\infty$, the expression (\ref{w1a+b+t}) goes into 
$\displaystyle W^{(1)}(a^{(2)}_+, b^{(1)}_+)\, =\,\Gamma_0
\times\,\sin^2 \frac{\theta_{ab}}{2}$, as it should.
For $M \to 0$,  $W^{(1)}(a^{(2)}_+, b^{(1)}_+, \tau) \to 0$ because
of the phase space reduction. 
For $\tau \to 0$ the expression (\ref{w1a+b+t}) has a pole for $\tau$:
\begin{eqnarray}
W^{(1)}(a^{(2)}_+, b^{(1)}_+, \tau \to 0)\, \approx \,\frac{\Gamma_0}{2}\,
\left (
\frac{1}{2}\, +\,\frac{2}{\pi (M\,\tau)}
\right )\,\times\,\sin^2 \frac{\theta_{ab}}{2}. \nonumber
\end{eqnarray}
The existence of the pole when $\tau \to 0$ has been discussed in many
works
\cite{Bernardini:1993qj,Joichi:1997xn,Maiani:1997pd,Joichi:1998hm}. It
was shown that the pole is not related to the ultraviolet divergence
of QFT, and, hence, can not be removed by usual
renormalisation technique. The presence of such ``surface'' divergencies
was noted long ago in \cite{bogolyubov}, where the concept of an
interaction in different regions of space-time was clearly stated.
The surface divergences are our ``penalty'' for using the Dirac picture
and the approximation of non-interaction in finite time.
However in our example, times $\tau$ are cut off by the time resolution
$\Delta t$ of the detector, which measures properties of the fermions
in the final state. It is obvious that $\Delta t \gg 1/M$. Hence the
pole $\tau$ в (\ref{w1a+b+t}) is not really essential here.

For the leading order to suffice, the full decay width should be much less
than the mass of the decaying particle. Let us suppose that the decay
$P \to f^+ f^-$ is absolutely dominant. Then the condition for using
perturbative theory is:
\begin{eqnarray}
\label{usl_primenimosti1}
\frac{\Gamma_0}{2}\, 
\left (
1\, +\, \frac{\textrm{si} (M\,\tau)}{\pi}\, +\, 
        \frac{\sin (M\,\tau)}{\pi (M\,\tau)^2}\, +\, 
        \frac{\cos (M\,\tau)}{\pi (M\,\tau)}
\right )\,
= \, W^{(1)}(\tau)\,\approx\,\Gamma (\tau)\,\ll\, M.
\end{eqnarray}
If the coupling constant $g$ is small enough, then it is possible to
obtain $\displaystyle\frac{\Gamma_0}{M} \ll 1$ as small as
needed. Hence condition (\ref{usl_primenimosti1}) is satisfied for
a wide range of the value $M\,\tau \gg 1$. 

Now let us return to the inequality (\ref{W-B-4}). Times $t_0$ and $t$
should be interpreted as the time intervals of measurement of each of
the probabilities. The probabilities of the flip of the spins of the
fermion $w\left (a^{(2)}_-(t_0) \to a^{(2)}_+ (t) \right )$ and
antifermion $w\left (b^{(1)}_-(t_0) \to b^{(1)}_+ (t) \right )$ in the
right part of (\ref{W-B-4}) are much smaller by the coupling constant
$g$. Hence at the leading order $w\left (a^{(2)}_-(t_0) \to a^{(2)}_+
  (t) \right ) \approx  w\left (b^{(1)}_-(t_0) \to b^{(1)}_+ (t)
\right ) \approx 0$. Analogously, without the spin flip $w\left
  (a^{(2)}_+(t_0) \to a^{(2)}_+ (t) \right ) \approx w\left
  (b^{(1)}_+(t_0) \to b^{(1)}_+ (t) \right ) \approx 1$.

Integration (\ref{w1a+b+t}) over $d \tau$ from $t_i$ to $t_f$ assuming
that $\displaystyle\frac{M}{\Gamma_0}\gg M\,\tau \gg 1$, gives:
\begin{eqnarray}
w\left (a^{(2)}_+, b^{(1)}_+, t_f - t_i \right ) \approx
\int\limits_{t_i}^{t_f} d\tau\, W^{(1)}(a^{(2)}_+, b^{(1)}_+,
\tau)\,\approx \\ \nonumber 
\approx\,
\frac{\Gamma_0}{2}\,\sin^2 \frac{\theta_{ab}}{2} \int\limits_{t_i}^{t_f} d\tau\, =\,
\frac{\left (t_f -t_i \right ) \Gamma_0}{2}\,\sin^2 \frac{\theta_{ab}}{2}.
\end{eqnarray}
Let the measurement time be $t_f - t_i = t_0$ on the right side of
(\ref{w1a+b+t}), and on the left side set it to $t_f - t_i = t$. Then:
\begin{eqnarray}
\label{bell-t/t_0}
\frac{t}{t_0}\,\sin^2\left (\frac{\theta_{ba}}{2} \right)\,\le\, 
\sin^2\left (\frac{\theta_{ca}}{2}\right )\, +\, 
\sin^2\left (\frac{\theta_{bc}}{2}\right ).
\end{eqnarray} 
This inequality differs from the classic inequality (\ref{W-B-2}) by
the ratio $\displaystyle\frac{t}{t_0} \ge 1$ on the left 
side. Note that it is correct to divide by $t_0$, because $t_0 \ge
\Delta t \gg 0$. The ratio $t/t_0$ may exceed unity by a few times,
meaning that the area of angular values which violate the inequality
is much wider. From the experimental point of view it is more
suitable to use the ratio of distances between the polarizers  $L/L_0$
instead of the ratio of times $t/t_0$. That is, in experiments measuring
the spin projections to ($\vec a$, $\vec c$) and ($\vec b$, $\vec c$),
it should be equal to $L_0$, and for $\vec a$ and $\vec b$ it should
be $L$, $\displaystyle\frac{M}{\Gamma_0}\gg M\, \{ L,\, L_0\} \gg 1$.

There are no terms in (\ref{bell-t/t_0}) specific to Hamiltonian
(\ref{Heff_for_PS2ff}), meaning this inequality is valid for any other
QFT Hamiltonian if it allows anticorrelation (\ref{pm=mp2}) in the
limits of applicability of the perturbation theory's leading order. In
this sense inequality (\ref{bell-t/t_0}) may be considered a 
universal time-dependent Bell inequality in the QFT in the absence 
of external fields.

%============================================================================================

\section{Conclusion}

Under the assumption of local realism and using Kolmogorov's
probability theory we obtain the time-dependent Bell inequalities in
Wigner form (\ref{W-B-4}). These inequalities may be used in
non-relativistic quantum mechanics as well as in the quantum field
theory, where it is impossible to exclude field interactions.
We consider a few examples to which the inequality (\ref{W-B-4}) may
be applied. In all cases we demonstrate the extension of the range of parameters
for the violation of (\ref{W-B-4}) in comparison to that of the classic
inequalities (\ref{W-B-2}). 

\section{Acknowledgements}
The authors would like to express our deep gratitude to Dr. S.~Baranov 
(Lebedev Physical Institute, Russia) for numerous discussions related 
to tests of the Bell inequalities in particle physics, to 
Dr. A.~Grinbaum (CEA-Saclay, SPEC/LARSIM, France) for educational 
chats on the foundations of quantum theory, to Prof. Sally Seidel 
(University of New Mexico, USA) for help with preparation of the 
paper, and to Aleister the Cat (Saint Genis-Pouilly, France) for happy 
purring, that inspired us through the time. 

This work was partially supported by the Russian Federation Scientific
Schools Support (grant number 3042.2014.2).

\end{document}